\documentclass[a4paper,conference]{IEEEtran}
\IEEEoverridecommandlockouts

\usepackage{cite}
\usepackage{amsmath,amssymb,amsfonts}
\usepackage{graphicx}
\usepackage{textcomp}
\usepackage{xcolor}
\usepackage[hidelinks,draft=false]{hyperref}
\usepackage[nolist]{acronym}
\usepackage{orcidlink}
\usepackage{mleftright}
\usepackage{siunitx}
\usepackage{tikz}
\usetikzlibrary{%
    angles,%
    arrows,%
    arrows.meta,%
    babel,%
    backgrounds,%
    calc,%
    fit,%
    matrix,%
    mindmap,%
    patterns,%
    positioning,%
    quotes,%
    shapes,%
    shadows,%
    tikzmark,%
    trees,%
}
\usepackage{pgfplots}
\usepgfplotslibrary{groupplots}
\usepgfplotslibrary{statistics}
\pgfplotsset{width=\textwidth*0.4,compat=1.9}
\usepackage{url}

\usepackage{algpseudocode}
\usepackage{algorithm}
\usepackage[libertine]{newtxmath}
\usepackage{bm}

\def\BibTeX{{\rm B\kern-.05em{\sc i\kern-.025em b}\kern-.08em
    T\kern-.1667em\lower.7ex\hbox{E}\kern-.125emX}}
\makeatletter
\def\endthebibliography{%
  \def\@noitemerr{\@latex@warning{Empty `thebibliography' environment}}%
  \endlist
}

\newtheorem{definition}{Definition}

\definecolor{kit-blue100}{cmyk}{.8,.5.,0,0}
\definecolor{kit-blue70}{cmyk}{.56,.35,0,0}
\definecolor{kit-blue50}{cmyk}{.4,.25,0,0}
\definecolor{kit-blue30}{cmyk}{.24,.15,0,0}
\definecolor{kit-blue15}{cmyk}{.12,.075,0,0}

\definecolor{kit-green100}{cmyk}{1,0,.6,0}
\definecolor{kit-green70}{cmyk}{.7,0,.42,0}
\definecolor{kit-green50}{cmyk}{.5,0,.3,0}
\definecolor{kit-green30}{cmyk}{.3,0,.18,0}
\definecolor{kit-green15}{cmyk}{.15,0,.09,0}

\definecolor{KITgreen}{rgb}{0,.59,.51}

\definecolor{KITpalegreen}{RGB}{130,190,60}
\colorlet{kit-maigreen100}{KITpalegreen}
\colorlet{kit-maigreen70}{KITpalegreen!70}
\colorlet{kit-maigreen50}{KITpalegreen!50}
\colorlet{kit-maigreen30}{KITpalegreen!30}
\colorlet{kit-maigreen15}{KITpalegreen!15}

\definecolor{KITblue}{rgb}{.27,.39,.66}

\definecolor{KITyellow}{rgb}{.98,.89,0}
\definecolor{kit-yellow100}{cmyk}{0,.05,1,0}
\definecolor{kit-yellow70}{cmyk}{0,.035,.7,0}
\definecolor{kit-yellow50}{cmyk}{0,.025,.5,0}
\definecolor{kit-yellow30}{cmyk}{0,.015,.3,0}
\definecolor{kit-yellow15}{cmyk}{0,.0075,.15,0}

\definecolor{KITorange}{rgb}{.87,.60,.10}
\definecolor{kit-orange100}{cmyk}{0,.45,1,0}
\definecolor{kit-orange70}{cmyk}{0,.315,.7,0}
\definecolor{kit-orange50}{cmyk}{0,.225,.5,0}
\definecolor{kit-orange30}{cmyk}{0,.135,.3,0}
\definecolor{kit-orange15}{cmyk}{0,.0675,.15,0}

\definecolor{KITred}{rgb}{.63,.13,.13}
\definecolor{kit-red100}{cmyk}{.25,1,1,0}
\definecolor{kit-red70}{cmyk}{.175,.7,.7,0}
\definecolor{kit-red50}{cmyk}{.125,.5,.5,0}
\definecolor{kit-red30}{cmyk}{.075,.3,.3,0}
\definecolor{kit-red15}{cmyk}{.0375,.15,.15,0}

\definecolor{KITpurple}{RGB}{160,0,120}
\colorlet{kit-purple100}{KITpurple}
\colorlet{kit-purple70}{KITpurple!70}
\colorlet{kit-purple50}{KITpurple!50}
\colorlet{kit-purple30}{KITpurple!30}
\colorlet{kit-purple15}{KITpurple!15}

\definecolor{KITcyanblue}{RGB}{80,170,230}
\colorlet{kit-cyanblue100}{KITcyanblue}
\colorlet{kit-cyanblue70}{KITcyanblue!70}
\colorlet{kit-cyanblue50}{KITcyanblue!50}
\colorlet{kit-cyanblue30}{KITcyanblue!30}
\colorlet{kit-cyanblue15}{KITcyanblue!15}

\begin{document}
    \title{A Deep Reinforcement Learning-based Approach for Adaptive Handover Protocols}
    \author{
        \IEEEauthorblockN{Johannes Voigt$^*$\, \orcidlink{0000-0002-4032-5577}, Peter J. Gu$^\dagger$\, \orcidlink{0009-0002-7853-9201}, and Peter M. Rost$^*$\, \orcidlink{0000-0002-8341-6989}}
        \IEEEauthorblockA{$^*$Communications Engineering Lab (CEL), Karlsruhe Institute of Technology (KIT), Germany}
        \IEEEauthorblockA{$^\dagger$Chair of Theoretical Information Technology, Technical University of Munich (TUM), Germany}
        Email: \texttt{johannes.voigt@kit.edu}, \texttt{peter.gu@tum.de}
    }
    
    \maketitle
    
    \begin{abstract}
        The use of higher frequencies in mobile communication systems leads to smaller cell sizes, resulting in the deployment of more base stations and an increase in handovers to support user mobility.
This can lead to frequent radio link failures and reduced data rates.
In this work, we propose a handover optimization method using proximal policy optimization (PPO) to develop an adaptive handover protocol.
Our PPO-based agent, implemented in the base stations, is highly adaptive to varying user equipment speeds and outperforms the 3GPP-standardized 5G NR handover procedure in terms of average data rate and radio link failure rate.
Additionally, our simulation environment is carefully designed to ensure high accuracy, realistic user movements, and fair benchmarking against the 3GPP handover method.

    \end{abstract}
    

    \section{Introduction}\label{sec:1_introduction}
The exponential growth in \ac{UE} connectivity and the rise of high-demand applications, such as autonomous driving, are increasing the operational complexity of cellular networks.
This complexity is further amplified by higher frequency bands, which enable faster transmission but lead to smaller cell sizes and dense \ac{HetNet} architectures.
These multi-tier networks, with a mix of macro and micro \acp{BS} require highly reliable and adaptive \ac{HO} protocols to maintain seamless connectivity.
Traditional network optimization methods based on static configurations and fixed parameters are insufficient to meet the dynamic needs of modern networks.

With increased user mobility and cell density, \ac{HO} events become more frequent, especially at cell boundaries where signal strength fluctuates.
This increase in \ac{HO} results in more \acp{PP} and \acp{RLF}, leading to service interruptions and rate losses.
\acp{PP}, occurring when \acp{UE} oscillate between neighboring cells, increase the \ac{HO} execution time and degrade \ac{QoS}.
The connectivity is further affected by \acp{RLF}, often due to too early or delayed \acp{HO}.
Thus, good timing of \acp{HO} is essential, but optimal protocol parameters vary widely depending on the network topology, \ac{UE} behavior, and mobility patterns, making it challenging to optimize them for all \acp{UE} in traditional event-based protocols.

To address these challenges, there is a need for protocols that can dynamically adapt to varying \ac{QoS} demands and environmental conditions.
\Ac{RL} represents a promising approach that enables protocol optimization based on feedback from the network.
\ac{RL}-based \ac{HO} protocols can optimize timing and \ac{HO} decision-making, adapting to \ac{UE} mobility and \ac{QoS} requirements.

\subsection{Related work}\label{sec:1-1-related-work}
The optimization of \ac{HO} protocols in cellular networks has been an active research area, especially with the shift towards dense \acp{HetNet} where high \ac{UE} mobility and smaller cell sizes increase the frequency of \acp{HO}.
Mobility management in LTE and 5G~NR, standardized by 3GPP, relies on an event-based protocol with fixed thresholds \cite{3gpp_ts_38_331}.
This method, however, is limited in adapting to varying conditions in modern networks, often leading to suboptimal performance \cite{tayyab_survey_2019}.

Adaptive \ac{HO} algorithms have been proposed to address these limitations. For instance, context-aware \ac{HO} mechanisms consider user speed to adjust \ac{HO} parameters, aiming to maximize \ac{UE} capacity \cite{guidolin_context-aware_2016, castro-hernandez_optimization_2018}.
These approaches rely on defined thresholds and are generally tailored for specific environments, limiting their adaptability.
Machine learning approaches offer more flexible solutions that adapt to changing network conditions.
Supervised learning has been used to predict optimal \acp{HO} based on historical data \cite{masri_machine-learning-based_2021}. However, these methods rely heavily on data quality and struggle to adapt to dynamic network changes.

In contrast, \ac{RL} provides a framework that learns by interacting with the environment, eliminating labeled datasets.
In \ac{HO} optimization, \ac{RL} algorithms show promise in reducing \acp{PP} and \acp{RLF} by learning adaptive policies that adjust \ac{HO} timing based on network conditions \cite{thillaigovindhan_comprehensive_2024}. Q-learning-based \ac{HO} methods have demonstrated gains by taking actions based on the \ac{RSRP} \cite{yajnanarayana_5g_2020}, but they face scalability issues in continuous state spaces.

\Ac{DRL} has led to more scalable solutions for \ac{HO} optimization. \ac{DRL} methods, such as deep Q-Networks and \ac{PPO}, have addressed \ac{HO} decisions in dense networks with complex decision spaces \cite{mollel_deep_2021, jang_proactive_2022}.
By using \acp{NN} to approximate optimal policies, these methods better handle variations in \ac{UE} mobility and network conditions.
\ac{DRL}-based \ac{HO} protocols reduce \acp{PP} and \acp{RLF} by optimizing \ac{HO} timing and cell selection based on \ac{QoS} requirements and mobility predictions.
Multi-agent \ac{DRL} approaches aim to optimize mobility network-wide but introduce higher complexity \cite{guo_joint_2020}.

\subsection{Contribution}\label{subsec:contribution}%
This work investigates the application of \ac{RL} in optimizing \ac{HO} protocols within cellular networks.
It aims to maximize the average data rate and reduce \acp{RLF} by enhancing the flexibility of \ac{HO} decision-making. Our approach optimizes the timing of \ac{HO} events, leading to higher \ac{SINR} during the \ac{HO} execution, and reducing the probability of \acp{RLF} due to \acp{HOF}.

Our results demonstrate that \ac{DRL}-based \ac{HO} protocols offer significant potential to improve the reliability, efficiency, and mobility-aware behavior required in next-generation cellular networks, especially in high-mobility environments.
The results are based on simulations modeling realistic \ac{UE} mobility patterns and network conditions, providing a comprehensive testbed for research on \ac{HO} optimization.
To support further research and reproducibility, we make the source code, datasets, and trained models publicly available\footnote{Source code, model, and datasets publicly available at: \url{https://github.com/kit-cel/HandoverOptimDRL}}.

    \section{Preliminaries}\label{sec:2_preliminaries}
\subsection{Handover Procedure within 5G NR}\label{subsec:ho_procedure}
Similar to previous 3GPP releases, mobility in 5G~NR is managed by an event-driven \ac{HO} procedure \cite{3gpp_ts_38_331}.
The procedure is based on \ac{RSRP} measurements of the serving cell and detected neighboring cells.
For intra-frequency \acp{HO}, as considered in this work, the \ac{HO} decision is usually based on \textit{Event A3}.
The event is triggered when the following condition is met, according to \cite[p. 260]{3gpp_ts_38_331}:
\begin{align}\label{eq:a3}
    M_\text{n} + \text{Off}_\text{n} + \text{Off}_\text{cn} - \text{Hys} &> M_\text{p} + \text{Off}_\text{p} + \text{Off}_\text{cp} + \text{Off}
    ,
\end{align}
where $M_\text{n}$ and $M_\text{p}$ are measurements of the neighboring and serving cell, neighboring cell-specific offsets $\text{Off}_\text{n}$, and $\text{Off}_\text{cn}$, serving cell offsets $\text{Off}_\text{p}$, $\text{Off}_\text{cp}$, $\text{Off}$, and a hysteresis \text{Hys}.

When Event A3 is triggered, the \ac{UE} starts sending measurement reports to the serving \ac{BS}, i.\,e., gNodeB, which then makes the \ac{HO} decision and initiates the following \ac{HO} preparation and finally sends the \ac{HO} command to the \ac{UE}, to start the \ac{HO} execution.
To avoid excessive \acp{HO}, the \ac{UE} only starts sending measurement reports after a certain waiting time, called \ac{TTT}.
If the \ac{RSRP} of the serving cell recovers before the \ac{HO} command is sent and exceeds a defined threshold, according to the leaving condition \cite[p. 260]{3gpp_ts_38_331}, the \ac{HO} can be aborted.
Too fast changes between the entering and leaving condition of Event A3 are reduced by the hysteresis in Eq.~\eqref{eq:a3}.
The offsets, \ac{TTT}, and the hysteresis are adjustable parameters configured by the serving \ac{BS}.
Network operators must optimize these parameters to prevent common \ac{HO} issues, such as \ac{RLF} caused by the \ac{UE} being out-of-synchronization due to poor \ac{SINR} or \ac{RSRP}, and \ac{PP} events, where the \ac{UE}  switches too often back to the previous \ac{BS} after a \ac{HO}.

\subsection{Background on Reinforcement Learning}\label{subsec:deep-rl}
In \ac{RL}, an agent interacts with an environment to maximize cumulative rewards over time.
At each time step $t$, the agent observes the state $s_t\in\mathcal{S}$, takes an action $a_t\in \mathcal{A}$ according to its policy $\pi$, which initiates the transition to the next state $s_{t+1}$ and receives a reward $r(s_t, a_t)=r_t$.
The policy $\pi(a\vert s)$ defines the probability of selecting action $a$ given state $s$.
In deep \acsu{RL}, the policy is represented by a \ac{NN}, called policy network, with parameters $\theta$, denoted as $\pi_\theta$.
The goal is to find an optimal policy $\pi^\star$ that maximizes the expected cumulative reward $G_t = \mathbb{E}_{\pi} \left[\sum_{k=0}^{\infty}\gamma^kr_{t+k+1}\right]$, where ${\gamma\in[0,1)}$ is the discount factor weighting future rewards.

Policy gradient methods are widely used to find the optimal policy $\pi^\star$.
The actor-critic framework  uses two separate \acp{NN}: an actor network, representing the policy $\pi_\theta$, and a critic network $v_\phi$, which estimates the expected return $G_t$ from state $s_t$, known as value function $V_t^\pi(s)$ \cite{sutton_reinforcement_2020}:
\begin{align}
    V^{\pi}_t(s)
    &=
    \mathbb{E}_{\pi} \mleft[\left.G_t \right\vert s_t \mright]
    =
    \mathbb{E}_{\pi}\mleft[\left.\sum_{k=0}^{\infty}\gamma^k r_{t+k+1}\right\vert s_t\mright] .
\end{align}

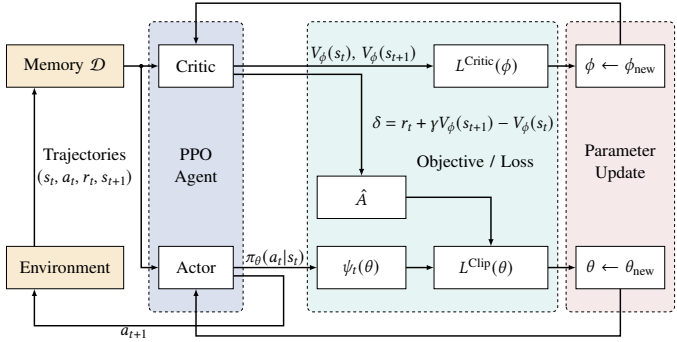
\begin{figure}[ht]
    \centering
    \tikzstyle{every picture}+=[remember picture]
\tikzset{>=latex}       %

\tikzstyle{arrow}   = [->,line width=1.5pt]
\tikzstyle{arrowr}  = [->]      %
\tikzstyle{arrowl}  = [<-]      %
\tikzstyle{arrowlr} = [<->]     %
\tikzstyle{doublearrow} = [arrow, double distance=1.75pt, arrows = {-Latex[length=0pt 2.5 0]}]

\tikzstyle{sum}         = [circle,draw,fill=white,node distance=1cm]
\tikzstyle{junction}    = [circle,fill,inner sep=3pt]
\tikzstyle{square}      = [rectangle,draw,minimum width=5cm,minimum height=4cm,text width=2cm,anchor=south west,align=center]
\tikzstyle{block}       = [rectangle,draw,minimum width=5cm,minimum height=3.0cm,text width=5cm,anchor=south west,align=center]
\tikzstyle{phantomblock}= [rectangle,minimum width=2cm,minimum height=1cm,text width=2cm,anchor=south west,align=center]

\tikzstyle{circles} = [circle,draw]

\tikzstyle{node}        = [circle,thick,draw=kit-blue100,minimum size=22,inner sep=0.5,outer sep=0.6]
\tikzstyle{node in}     = [node,green!20!black,draw=kit-green100!30!black,fill=kit-green100!25]
\tikzstyle{node hidden} = [node,blue!20!black,draw=kit-blue100!30!black,fill=kit-blue100!20]
\tikzstyle{node convol} = [node,orange!20!black,draw=kit-orange100!30!black,fill=kit-orange100!20]
\tikzstyle{node out}    = [node,red!20!black,draw=kit-red100!30!black,fill=kit-red100!20]
\tikzstyle{connect}     = [thick,black]
\tikzstyle{connect arrow}=[-{Latex[length=4,width=3.5]},thick,kit-blue100,shorten <=0.5,shorten >=1]
\tikzset{ %
  node 1/.style={node in},
  node 2/.style={node hidden},
  node 3/.style={node out},
}
\def\nstyle{int(\lay<\Nnodlen?min(2,\lay):3)} %

\def\linestyles{{%
    "solid",%
    "loosely dashed",%
    "dashed",%
    "densely dashed",%
    "loosely dotted",%
    "dotted",%
    "densely dotted",%
    "loosely dash dot",%
    "dash dot",%
    "densely dash dot",%
    "loosely dash dot dot",%
    "dash dot dot",%
    "densely dash dot dot"%
}}
\resizebox{0.49\textwidth}{!}{
\begin{tikzpicture}[font=\fontsize{35}{45}\selectfont, transform shape,every node/.style={scale=0.51, align=center}]
    \node[phantomblock,minimum width=5cm,text width=5cm](middle){$\begin{matrix}\text{PPO}\\\text{Agent}\end{matrix}$};
    \node[block,above=2cm of middle,fill=white](critic){Critic};
    \node[block,below=2cm of middle,fill=white](actor){Actor};

    \node[block,left=1.25cm of critic,minimum width=8cm,text width=8cm,fill=KITorange!20](memory){Memory $\mathcal{D}$};
    \node[block,left=1.25cm of actor,fill=KITorange!20,minimum width=8cm,text width=8cm](env){Environment};
    \draw[arrow]($(env.north)-(1.1,0)$)--($(memory.south)-(1.1,0)$)node[midway,right]{Trajectories\\$(s_t,\hspace{0.05em}a_t,\hspace{0.05em}r_t,\hspace{0.05em}s_{t+1})$};
    \draw[arrow](memory)--(critic)node[junction,midway](mid_exp_cr){};
    \draw[arrow](mid_exp_cr.center)|-(actor){};
    \draw[arrow]($(actor.east)-(0,.3)$)-|($(actor.east)-(-1.8,2.1)$)-|($(env.south)-(1.1,0)$)node[below,pos=0.3]{$a_{t+1}$};

    \node[block,right=3cm of actor,minimum width=6cm,text width=6cm,fill=white](prob){$\psi_t(\theta)$};
    \node[block,above=of prob,minimum width=6cm,text width=6cm,fill=white](advantage){$\hat{A}$};
    \draw[arrow](actor)--(prob)node[above,pos=.5]{$\pi_\theta(a_t|s_t)$};

    \node[block,right=1cm of prob,fill=white,minimum width=8cm](lclip){$L^{\text{Clip}}(\theta)\,\,\,$};
    \draw[arrow](prob)--(lclip){};
    \draw[arrow](advantage)-|(lclip){};
    \node[block,fill=white,minimum width=8cm,align=left]at($(critic.south-|lclip.west)$)(lcritic){{$L^{\text{Critic}}(\phi)$}};
    \draw[arrow](critic)--(lcritic)node[above,pos=0.65]{$V_\phi(s_t),\,V_\phi(s_{t+1})$};
    \draw[arrow]($(critic.east)-(0,0.3)$)-|(advantage)node[right,pos=.75]{$\,\,\,\delta=r_t + \gamma V_\phi(s_{t+1}) - V_\phi(s_{t})$};

    \node[block, right=1cm of lcritic,fill=white,minimum width=6cm,text width=6cm](sgd1){$\phi \leftarrow \phi_\text{new}$};
    \node[block, right=1cm of lclip,fill=white,minimum width=6cm,text width=6cm](sgd2){$\theta \leftarrow \theta_\text{new}$};
    \draw[arrow](lcritic)--(sgd1){};
    \draw[arrow](lclip)--(sgd2){};
    \draw[arrow](sgd1.north)--($(sgd1.north)+(0,1.5)$)-|(critic){};
    \node[phantomblock,below=2.0cm of sgd1,minimum width=5cm,text width=5cm](optim){Parameter\\Update};
    \draw[arrow](sgd2.south)--($(sgd2.south)-(0,1.7)$)-|(actor){};

    \node[phantomblock,left=1.5cm of optim,minimum width=9cm,text width=9cm](){Objective / Loss};

    \begin{scope}[on background layer]
        \draw[dashed,rounded corners=2mm,fill=KITblue!20]($(actor.west)-(0.35,1.6)$)rectangle($(critic.east)+(0.35,1.6)$);
        \draw[dashed,rounded corners=2mm,fill=KITgreen!10]($(prob.west)-(0.35,1.6)$)rectangle($(lcritic.east)+(0.35,1.6)$);
        \draw[dashed,rounded corners=2mm,fill=KITred!10]($(sgd2.west)-(0.35,1.6)$)rectangle($(sgd1.east)+(0.35,1.6)$);
    \end{scope}
\end{tikzpicture}
}
    \caption{The centralized \ac{DRL} agent includes the environment and trajectory buffer (orange), a proximal policy optimization actor-critic unit (blue), objective, i.\,e., loss computation (green), and an optimizer to update the network parameters (red).}
    \label{fig:ppo}
\end{figure}

Policy gradient methods directly optimize the policy by adjusting parameters $\theta$ through gradient ascent on the objective function.
The gradient with respect to $\theta$ can be estimated as:
\begin{align}
    \hat{g} &= \hat{\mathbb{E}}_t \mleft[ \nabla_{\theta} \log \pi_{\theta}(a_t | s_t) \hat{A}_t \mright]
    ,
\end{align}
where $\hat{A}_t$ is an estimate of the advantage function based on $V_\phi$ \cite{schulman_proximal_2017}. The \acl{GAE} \cite{schulman_high-dimensional_2018} can be used to estimate $\hat{A}_t$, see Fig~\ref{fig:ppo}.

\ac{PPO} improves the performance by constraining policy updates to prevent drastic changes \cite{schulman_proximal_2017}.
This is achieved by optimizing a truncated, i.\,e., proximal, objective function and thereby improves training stability:
\begin{align}
    L^{\text{Clip}}(\theta) = \hat{\mathbb{E}}_t \left[ \min \left( \psi_t(\theta) \hat{A}_t, ~\text{clip}(\psi_t(\theta), 1 - \epsilon, 1 + \epsilon) \hat{A}_t \right) \right],
\end{align}
where $\psi_t(\theta) = \frac{\pi_{\theta}(a_t | s_t)}{\pi_{\theta_{\text{old}}}(a_t | s_t)}$ is the probability ratio between new and old policies, and $\epsilon$ is a hyperparameter that limits policy updates.
The critic minimizes the mean squared error between the estimated value $V_\phi(s)$ and a target value ${V_\text{target} =r_t+\gamma V_\phi(s_{t+1})}$ \cite{schulman_proximal_2017}, as shown in Fig.~\ref{fig:ppo}:
\begin{align}
    L^\text{Critic}(\phi)=\mathbb{E}_t\mleft[(V_\phi(s_t)-V_\text{target})^2\mright]
    .
\end{align}
The policy and value network are optimized using Adam \cite{kingma_adam_2017} with learning rate $\alpha$.

    \section{System Model}\label{sec:3_system_model}
\subsection{The Environment}\label{subsec:system_model}
The considered scenario is a dense urban 5G NR network with a set of $N$ macro \acp{BS}, denoted as $\mathcal{B}=\{b_0, \ldots, b_{N-1}\}$, and \acp{UE} moving randomly through the simulated area at different speeds.
The \acp{BS} positions are based on real locations from publicly available network information, as shown in Fig.~\ref{fig:karlsruhe_map}.
All \acp{BS} and \acp{UE} use the IMT-2000 (n1) band at $\SI{2.1}{GHz}$ for data transmission.

We use the \textit{Vienna 5G System Level Simulator}~\cite{Vienna5GSLS} with environmental data such as building floor plans and street systems obtained by OpenStreetMap.
The region of interest\footnote{Coordinates: 4\ang 900'17.6"N \ang 822'38.3"E -- 49°00'40.3"N 8°23'44.2"E} is near the city center and includes pedestrian zones, side roads, and main streets covering diverse mobility scenarios. \ac{UE} speeds range from $\SI{3}{km/h}$ to $\SI{50}{km/h}$, representing pedestrians, cyclists, and cars.
To test adaptability, HO protocols in Sec.~\ref{sec:5_results} will also be evaluated at higher speeds up to $\SI{90}{km/h}$.

To generate realistic movement patterns and keep users on roads, we use \textit{\acl{SUMO}}~\cite{SUMO2018}.
GPX tracks for \acp{UE} are sampled every $\Delta t=\SI{10}{ms}$.
The temporal resolution $\Delta t$ is chosen such that handover-related procedures defined in Sec.~\ref{subsec:ho_procedure} and Sec.~\ref{subsec:sim_ho_rlf_pp} can be correctly resolved and simulated.
These tracks are then used within the Vienna 5G Simulator to calculate the \ac{RSRP} and \ac{SINR} values for each \ac{UE}-\ac{BS} connection at each time step, based on the UMa~5G channel model \cite{3gpp_tr_38_901}.
Line-of-sight and shadowing effects are derived from 3D building data.
The generated \ac{RSRP} and \ac{SINR} are assumed to be layer 1 filtered to reduce fast-fading effects. Additionally, layer 3 filtering is applied according to \cite{3gpp_ts_38_331}.

\subsection{Simulation of Handover, PP, and RLF Events}\label{subsec:sim_ho_rlf_pp}
Accurately simulating \acp{HO}, \acp{PP}, and \acp{RLF} is essential for comparing 3GPP mobility management with the \ac{DRL}-based approach.
\acp{HO} are modeled with a preparation phase followed by an execution phase.
During preparation, \acp{HO} can be aborted if signal conditions change.
The \ac{HO} command is received by the \ac{UE} at the end of the \ac{HO} preparation, after which the \ac{UE} connects to the target \ac{BS}.
The following definitions, based on \cite{3gpp_tr_36_839}, are used to represent \acp{RLF}, \acp{HOF}, and \acp{PP} in the simulation:

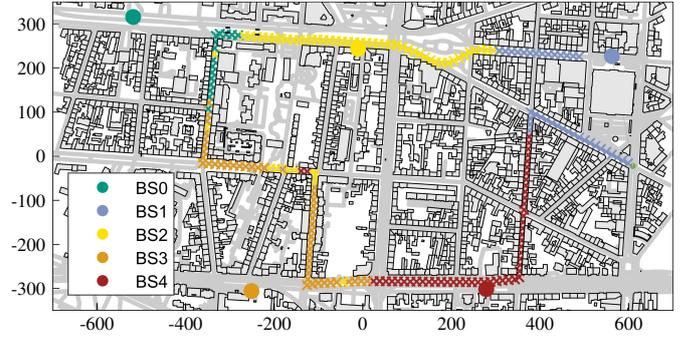
\begin{figure}[ht]
    \centering
    \begin{tikzpicture}[x=0.45\textwidth/14,y=0.45\textwidth/14]
        \node[anchor=center,inner sep=0] at (0,0)(fig){\includegraphics[width=0.45\textwidth]{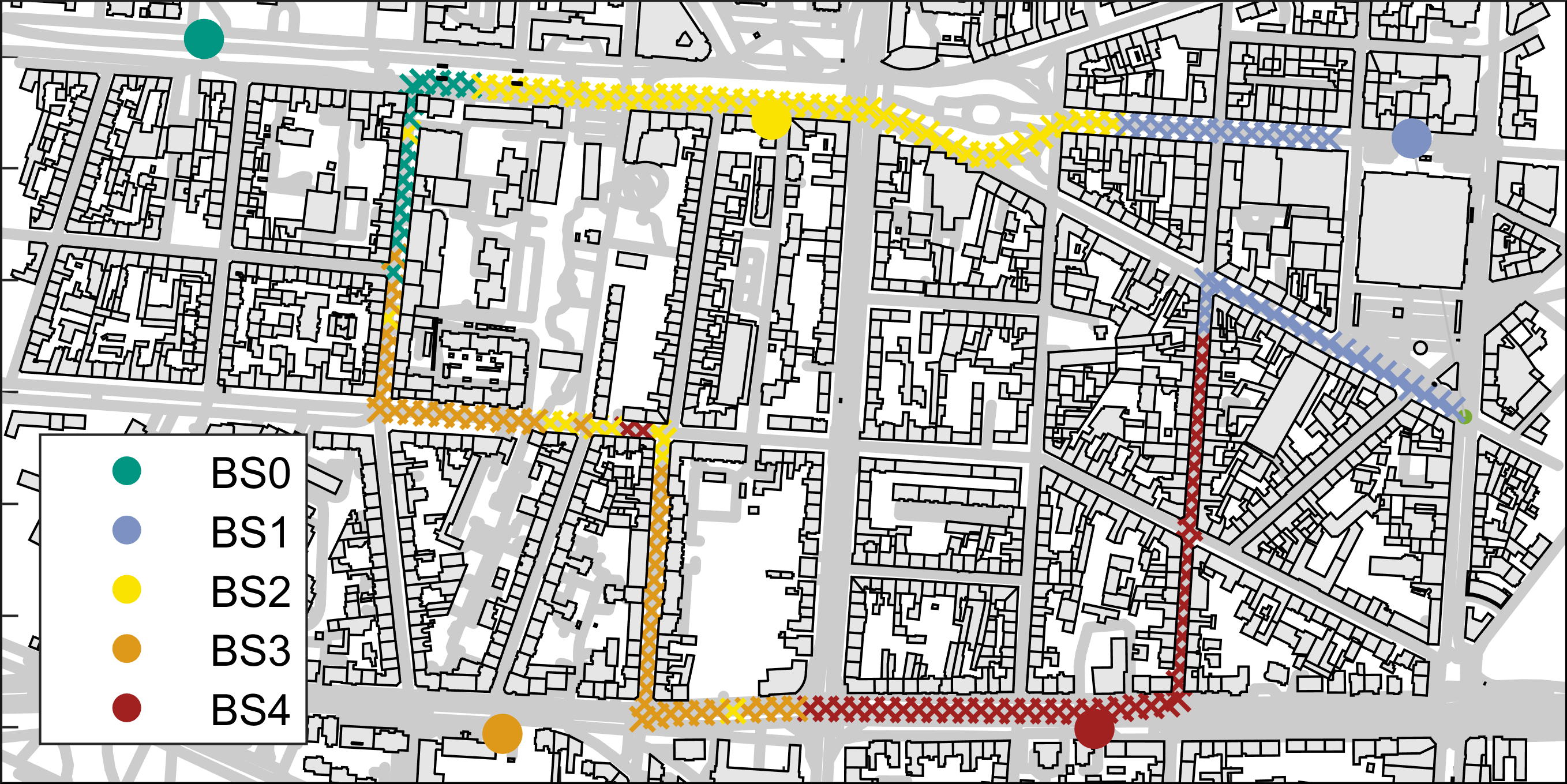}};
        \begin{scope}[xscale=1, yscale=1]
            \foreach \x in {-600,-400,...,600} {
                \draw[line width=0.4] ({\x/100}, -3.335) -- ({\x/100}, -3.49);
            }
            \foreach \x in {-600,-400,...,600} {
                \node[below]at({\x/100}, -3.45){\scriptsize\x};
            }
            
            \foreach \y in {-300,-200,...,300} {
                \node[left]at(-6.95, {\y/100}) {\scriptsize\y};
            }
        \end{scope}
    \end{tikzpicture}
    \caption{Simulated region with areas of different speed limits and pedestrian zones. \acp{BS} are indicated by circles and crosses mark the path of an \ac{UE} in the color of the serving \ac{BS}. The axes indicate the distance in meters.}   
    \label{fig:karlsruhe_map}
\end{figure}

\begin{definition}[Radio Link Failure]
    An \ac{RLF} occurs if the \ac{UE} experiences out-of-sync conditions, i.\.e., $\text{SINR} < Q_\text{out}$. After N310 consecutive out-of-sync indicators from lower layers, timer T310 will be started. If T310 expires before recovery, i.\,e., N311 in-sync indicators ($\text{SINR}>Q_\text{in}$) are received, an \ac{RLF} is declared.
\end{definition}

\begin{figure}[hb]
    \centering
    \tikzstyle{process} = [rectangle, 
minimum width=2cm, 
minimum height=1cm, 
text centered, 
text width=2cm, 
draw=black, 
]
\tikzstyle{process2} = [rectangle, 
minimum width=0cm, 
minimum height=0cm, 
text centered, 
text width=0cm, 
draw=white, 
]
\tikzstyle{process3} = [rectangle, 
minimum width=5cm, 
minimum height=1cm, 
text centered, 
text width=4cm, 
draw=black, 
]
\tikzstyle{process4} = [rectangle, 
minimum width=3cm, 
minimum height=1cm, 
text width=2cm, 
draw=black, 
]
\tikzstyle{arrow} = [draw=red,thick,->,>=latex]
\tikzstyle{arrow2} = [draw=black,thick,->,>=latex]	
\tikzstyle{arrow3} = [draw=black,thick,<->,>=latex]

\scalebox{.8}{\begin{tikzpicture}[node distance=2cm]
    \node (start) [process2]{};
    \node (in1) [process, right of=start, xshift=1.62cm] {HO prep.};
    \node (pro1) [process, right of=in1, xshift=0.23cm] {HO exe.};
    \node (optimize) [process3, above = 1cm of in1, xshift=-1.385cm] {T310};
    \node (optimize2) [process4, right of =optimize, xshift=2cm] {\hspace{1.5em}T310};
    \draw [arrow] (4.73,1.5) -- node[anchor=east] {\textcolor{red}{HOF}}(4.73,0.5);
    \draw[black] (5.7,1) to (6.7,3);
    \draw [arrow2] (4.73,3.5) -- node[yshift = 1.1em, anchor=south] {reset T310}(4.73,2.5);
    \draw [arrow2] (-0.27,3.5) -- node[yshift = 1.1em, anchor=south] {SINR$<Q_{\text{out}}$}(-0.27,2.5);
    \draw [arrow3] (-0.27,3) -- node[anchor=south] {SINR$<Q_{\text{in}}$}(4.73,3);
\end{tikzpicture}}
    \caption{\ac{HOF} triggered after \ac{HO} preparation, causing a \ac{RLF}.}
    \label{fig:hof}
\end{figure}
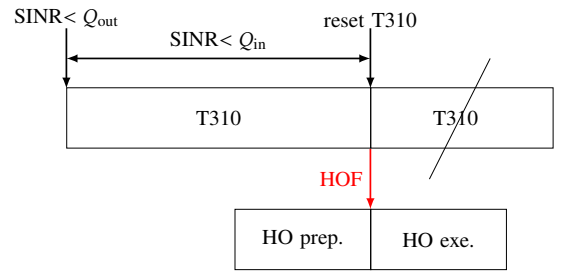

\begin{definition}[Handover Failure]
    An \ac{HOF} triggers an \ac{RLF} if the \ac{HO} command is received while timer T310 is running, as shown in Fig.~\ref{fig:hof}. Additionally, if T310 expires before the \ac{HO} execution starts, an \ac{HOF} and \ac{RLF} occurs.
\end{definition}

\begin{definition}[Ping-pong Handover]
    A \ac{PP} is defined as a \ac{HO} where the \ac{UE} switches to a new \ac{BS} and then returns to the previous \ac{BS} within a period shorter than a specified \ac{MTS}.
\end{definition}
Table~\ref{tab:3gpp} shows the parameters for modeling \acp{PP}, \acp{RLF}, and the 3GPP event-based \ac{HO} procedure \cite{castro-hernandez_optimization_2018}\cite{3gpp_tr_36_839}.

\begin{table}[ht]
    \caption{Parameters used for 3GPP protocol \& \ac{RLF} detection}
    \centering
    \fontsize{10pt}{14pt}\selectfont
    \begin{tabular}{{l}{c}}
    \hline
    Parameter & Value \\
    \hline 
    $Q_{\mathrm{in}}$ & \SI{-6}{dB} \\
    $Q_{\mathrm{out}}$ & \SI{-8}{dB} \\
    A3 hysteresis ($\text{Hys}$)& \SI{1}{dB} \\
    A3 Offset ($\text{Off}$)& [0, 1, 2]\,\SI{}{dB} \\
    A3 \ac{TTT} & [40, 80, 160]\,\SI{}{ms} \\
    \ac{HO} preparation time & \SI{50}{ms} \\
    \ac{HO} execution time & \SI{40}{ms} \\
    MTS & \SI{1000}{ms}\\
    T310 & \SI{1000}{ms} \\
    Avg. \ac{RLF} recovery & \SI{200}{ms} \\
    Counter N310 & 10\\
    Counter N311 & 3\\
    \hline
    \end{tabular}
    \label{tab:3gpp}
\end{table}

    \section{RL Problem Formulation and Objective}\label{sec:4_ho_optim}
The goal of the \ac{PPO}-based agent is to maximize the average rate of the \acp{UE}.
We model the problem based on Sec.~\ref{sec:2_preliminaries}, such that the agent is localized in each \ac{BS} and receives measurements from the \acp{UE} on which the agent's decisions are based.
It is assumed that the agent is trained in advance (before deployment) to be ready for operation, but it can be further optimized dynamically by updating the network parameters.

\subsection{State}
The state at any time $t$, denoted by $\bm{s}_t$, is represented by a vector containing important information about the network conditions and the status of the \ac{UE}.
The state $\bm{s}_t$ includes a one-hot encoded vector $\bm{s}_t^\text{BS} = \begin{bmatrix} 0,\, \dots,\, 1,\, \dots,\, 0 \end{bmatrix} \in\mathbb{F}_2^N$ to indicate the current serving \ac{BS}, where the position corresponding to the serving BS among $N$ available \acp{BS} is $1$, while all other entries are $0$, allowing the agent to identify the serving \ac{BS}.
The $i$-th entry of $\bm{s}_t$, denoted $\bm{s}_t(i)$, is defined as follows:
\begin{align}
    \bm{s}_t^\text{BS}(i) &= \begin{cases} 
        1, & \text{if $b_i \in\mathcal{B}$ is the serving BS}\\ 
        0, & \text{otherwise}.
    \end{cases}
\end{align}

To add information about the signal quality from each \ac{BS}, we include the \ac{SINR}.
The \ac{SINR} is clipped to $[-10, 10]$\,dB according to the \ac{ECDF} in the time period around the handover to focus on the \ac{HO} decision-making.
This speeds up training because there is no need for \acp{HO} during phases where the serving BS is clearly the best BS.
The clipped \ac{SINR} values are then scaled to $[0, 1]$, providing a normalized indication of connectivity quality in the same scale as the one-hot encoded \ac{BS} indicator vector $\bm{s}_t^\text{BS}$:
\begin{align}
    \bm{s}_t^\text{SINR} &= \begin{bmatrix} q_0, q_1, \ldots, q_{N-1} \end{bmatrix},
\end{align}
where each $q_i \in [0, 1]$ represents the clipped and scaled \ac{SINR} of BS $b_i\in\mathcal{B}$. This vector helps the agent evaluate the relative connection quality across potential target \acp{BS} for \acp{HO}.

Furthermore, to discourage \acp{PP}, the state vector includes a binary indicator flag. This indicator is set to $1$ if the elapsed time since the last \ac{HO}, $t - t_\text{HO}$, is less than a specified \ac{MTS}, and $0$ otherwise:
\begin{align}
    s_t^\text{PP} &= \begin{cases} 
        1, & \text{if } t - t_\text{HO} < \text{MTS} \\ 
        0, & \text{otherwise},
    \end{cases}
\end{align}
where $t_\text{HO}$ denotes the last \ac{HO} time. Including this \ac{PP} indicator aids the agent in stabilizing the \ac{UE} connection by avoiding oscillations between \acp{BS}.
Thus, the full state vector $\bm{s}_t$ at time $t$ and of size $2N+a$ is constructed as:
\begin{align}
    \bm{s}_t &= \begin{bmatrix} \bm{s}_t^\text{BS}, \bm{s}_t^\text{SINR}, s_t^\text{PP} \end{bmatrix} \in \mathcal{S}.
\end{align}

\subsection{Action}
At each time step $t$, the agent chooses an action ${a_t \in \mathcal{A} = \{0, \ldots, N-1\}}$ that corresponds to the index of a \ac{BS}. The action either initiates a \ac{HO} to a neighboring \ac{BS} or the \ac{UE} remains connected to the serving \ac{BS}.
If the agent selects a different target \ac{BS} while a \ac{HO} is already in preparation, the current preparation is aborted, the target \ac{BS} is updated, and the \ac{HO} process restarts.

\subsection{Reward}
The reward function $r_t(\bm{s}_t, a_t)$ is designed to encourage stable, high-quality connections and to avoid \acp{PP} and \acp{RLF} by penalizing them.
The reward consists of three components:
\begin{align*}
    r_t(\bm{s}_t, a_t) &= r_\text{SINR}(\bm{s}_t, a_t) + r_\text{PP}(\bm{s}_t, a_t) + r_\text{RLF}(\bm{s}_t, a_t).
\end{align*}
Each component relates to a specific aspect of \ac{HO} performance, where $r_\text{SINR}(\bm{s}_t, a_t)$ rewarding high \ac{SINR} values and encouraging the agent to select \acp{BS} that provide better connectivity, with a bonus $C\in\mathbb{R}$ for the currently best \ac{BS}:
\begin{align}
    r_\text{SINR}(\bm{s}_t, a_t) &= \begin{cases}
        \bm{s}_t^\text{SINR}(i) + C, & \text{if } i = \arg\max_{i \in \mathcal{B}} \bm{s}_t^\text{SINR}(i) \\ 
        \bm{s}_t^\text{SINR}(i), & \text{otherwise}.
    \end{cases}
\end{align}
To prevent \acp{PP}, the second component, $r_\text{PP}(\bm{s}_t, a_t)$, applies a penalty when a \ac{PP} occurs:
\begin{align}
    r_\text{PP}(\bm{s}_t, a_t) &= \begin{cases}
        -C, & \text{if PP detected} \\ 
        0, & \text{otherwise}.
    \end{cases}
\end{align}
Similar to the predecessor, $r_\text{RLF}(\bm{s}_t, a_t)$ penalizes the agent for allowing the \ac{UE} to experience out-of-sync conditions or \acp{RLF}, emphasizing the importance of avoiding service disruptions:
\begin{align}
    r_\text{RLF}(\bm{s}_t, a_t) &= \begin{cases}
        -C, & \text{if SINR indicates out-of-sync} \\ 
        -2C, & \text{if RLF detected} \\ 
        0, & \text{otherwise}.
    \end{cases}
\end{align}

\subsection{Objective}\label{subsec:problem_formulation}
In this framework, the \ac{PPO} agent aims to maximize the average data rate by learning an optimal policy for \ac{HO} decisions.
The average rate $\overline{R}$ is calculated as:
\begin{align}
    \overline{R} &= \frac{1}{T} \sum_{t=0}^{T-1} B  \log_2\mleft(1 + \text{SINR}_{\bm{b}(t)}(t)\mright),
\end{align}
where $B$ represents the bandwidth, $T$ is the total number of time steps, and $\text{SINR}_{\bm{b}(t)}(t)$ is the \ac{SINR} at the serving \ac{BS} $\bm{b}(t) \in \mathcal{B}$ at time $t$.

To provide an upper bound for the achievable rate, we define $\overline{R}_\text{max}$ as the maximum possible data rate if the \ac{HO} execution time was zero, allowing the \ac{UE} to always connect to the BS with the highest \ac{SINR}. This idealized rate is given by:
\begin{align}
    \overline{R}_\text{max} &= \frac{1}{T} \sum_{t=0}^{T-1} \max_{b \in \mathcal{B}} B \log_2\mleft(1 + \text{SINR}_{b}(t)\mright).
\end{align}
The agent’s performance is then assessed through the relative average rate $\Gamma_\text{R} \in[0,1]$, defined as:
\begin{align}
    \Gamma_\text{R} &= \frac{\overline{R}}{\overline{R}_\text{max}}\,.
    \label{eq:gamma-r}
\end{align}
This metric, $\Gamma_\text{R}$, allows us to evaluate how closely the agent’s policy approximates the optimal rate in the absence of \ac{HO} delays and to compare it with the 3GPP \ac{HO} procedure as a benchmark.

In addition to maximizing data rate, a secondary objective is to minimize \acp{RLF} by minimizing \acp{HOF}, thereby reducing service interruptions and improving connection stability. By balancing these objectives, the agent seeks to optimize both the quality and reliability of the UE's connection within the network.

\subsection{Training Process}
In each episode, a random training dataset is selected to support training. To further enhance generalization and prevent the agent from becoming reliant on a specific \ac{BS}, the \ac{BS} indices $b\in\mathcal{B}$, are shuffled after each episode.
The training process consists of two major phases. In the first phase, the episode terminates only if a \ac{RLF} is detected or the maximum number of episode time steps $T$ is reached, allowing the agent to focus on learning optimal \ac{HO} strategies.
In the second phase, episodes are also terminated by a \ac{PP} event, introducing a penalty for excessive \acp{HO} and encouraging more stable connections.

The agent is implemented using \ac{PPO} from the Stable-Baselines3 library \cite{stable-baselines3}. Tab.~\ref{tab:ppo} provides training-specific parameters determined by hyperparameter optimization.

\begin{table}[ht]
    \caption{Training \& \ac{PPO} agent related parameters}
    \centering
    \fontsize{10pt}{14pt}\selectfont
    \begin{tabular}{{l}{c}}
        \hline
        Parameter & Value  \\
        \hline
        C (reward/penalty) & $0.95$ \\
        Learning rate $\alpha$ & $5\cdot10^{-5}$\\
        \ac{NN} architecture & $[64, 128, 64]$\\ 
        PPO $\mathrm{ent}_{\mathrm{coef}}$ & 0.1 \\
        \hline
    \end{tabular}
    \label{tab:ppo}
\end{table}

    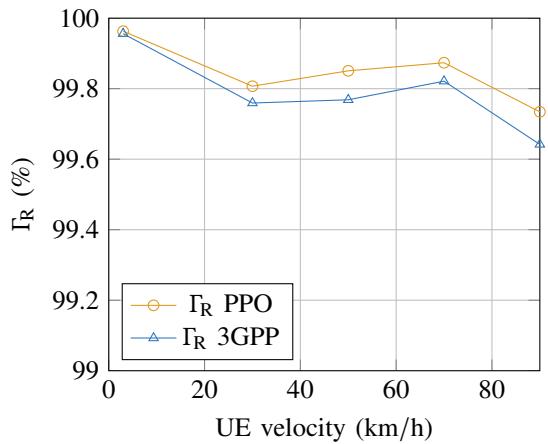
\begin{figure}[ht]
    \centering
    \begin{tikzpicture}
        \begin{axis}[
            xlabel={UE velocity ($\SI{}{km/h}$}),
            ylabel = $\Gamma_\text{R}$ (\%),
            ymin = 99,
            ymax = 100,
            xmin = 0,
            xmax = 90,
            grid=both,
            legend pos=south west,
            ]
            \addplot[mark=o,KITorange] coordinates { %
            (3, 99.963)
            (30,99.80695183431691)
            (50,99.85057001324993)
            (70,99.87354793794199)
            (90,99.73432937576673)
            };
            \addplot[mark=triangle, kit-blue100] coordinates { %
            ( 3,99.956)
            (30,99.75925946499762)
            (50,99.7684872148118)
            (70,99.82122371780747)
            (90,99.64192179179121)
            }; 
            \addlegendentry{$\Gamma_\text{R}$ PPO}
            \addlegendentry{$\Gamma_\text{R}$ 3GPP}
        \end{axis}
    \end{tikzpicture}
    \caption{Achieved relative average rate $\Gamma_\text{R}$ of the 3GPP and PPO-based \ac{HO} protocol for different \ac{UE} velocities.}
    \label{fig:datarate}
\end{figure}

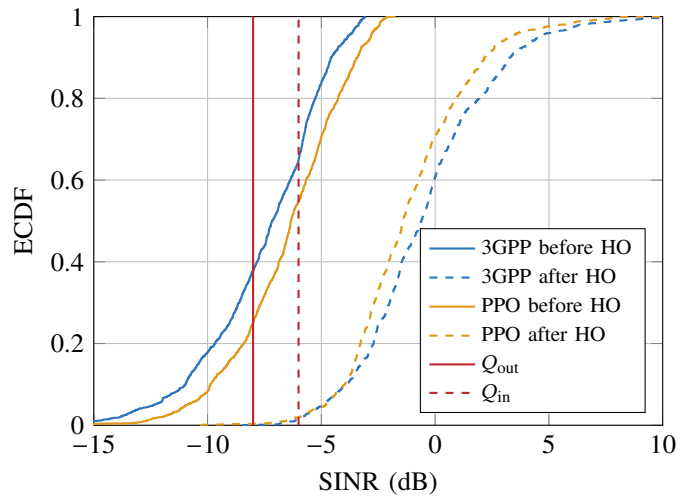
\begin{figure}[hb]
\begin{tikzpicture}
    \begin{axis}[
        width=0.5\textwidth,
        height=7cm,
        xlabel={SINR ($\SI{}{dB}$)},
        ylabel={ECDF},
        xmin=-15, xmax=10,
        ymin=0, ymax=1,
        grid=both,
        legend cell align={left},
        legend pos=south east,
        legend style={font=\footnotesize},
    ]
    
    \addplot[kit-blue100, solid, thick] table [x index=0, y index=1, col sep=comma] {figures/ecdf/3gpp_pcell.csv};
    \addlegendentry{3GPP before HO}

    \addplot[kit-blue100, dashed, thick] table [x index=0, y index=1, col sep=comma] {figures/ecdf/3gpp_ncell.csv};
    \addlegendentry{3GPP after HO}
    
    \addplot[KITorange, solid, thick] table [x index=0, y index=1, col sep=comma] {figures/ecdf/ppo_pcell.csv};
    \addlegendentry{PPO before HO}

    \addplot[KITorange, dashed, thick] table [x index=0, y index=1, col sep=comma] {figures/ecdf/ppo_ncell.csv};
    \addlegendentry{PPO after HO}

    \addplot[kit-red100, thick] coordinates {(-8,0) (-8,1)};
    \addlegendentry{$Q_\text{out}$}

    \addplot[kit-red100, dashed, thick] coordinates {(-6,0) (-6,1)};
    \addlegendentry{$Q_\text{in}$}

    \end{axis}
\end{tikzpicture}
\caption{ECDF of SINR before (solid) and after (dashed) handover execution under 3GPP and PPO policies. Red lines indicate $Q_\text{out}$ (solid) and $Q_\text{in}$ (dashed) SINR \ac{QoS} thresholds.}
\label{fig:ecdf}
\end{figure}

\section{Numerical Results}\label{sec:5_results}
The performance of both the 3GPP \ac{HO} procedure and the \acl{PPO} agent was evaluated on datasets generated using the setup described in Sec.~\ref{sec:3_system_model}.
\acp{UE} moving at speeds of up to $\SI{50}{km/h}$ were used, with parameters provided in Tab.~\ref{tab:3gpp} and Tab.~\ref{tab:ppo} to optimize the parameters of the 3GPP \ac{HO} procedure and to train the \ac{PPO} agent.
The 3GPP \ac{HO} protocol was optimized to perform well on average on the training data sets.
However, no additional optimizations for individual \ac{UE} speeds were made for either protocol, as the adaptivity of the protocols is being investigated.

Both protocols were tested on randomly generated datasets, with additional tests using \acp{UE} traveling at speeds up to $\SI{90}{km/h}$ to evaluate adaptability to varying environmental conditions.
During testing, the \ac{PPO} agent was not terminated upon experiencing \ac{PP} events or during the \ac{RLF} recovery.
Actions during these periods are ignored because the agent cannot perform actions during this time, i.\,e., the actions do not change the state of the environment.
However, as previously described, the \ac{HO} preparation could be modified or aborted if necessary.

Fig.~\ref{fig:datarate} presents the relative average rate $\Gamma_\text{R}$ as calculated in Eq.~\ref{eq:gamma-r} for both the reference protocol and the \ac{PPO} agent across varying \ac{UE} speeds, averaged over all test datasets.
The results show that our \ac{PPO}-based approach achieves slightly better relative average rates than the reference protocol.
In particular, the performance gap to the theoretical upper bound remains relatively small.
This is mostly because \acp{UE} are connected to the optimal \ac{BS} most of the time, and the relative duration required for \ac{HO} execution or \ac{RLF} recovery is minimal.
However, as \ac{UE} speeds increase, this gap also increases, which is expected since higher speeds require more frequent \acp{HO}, where precise timing is more critical.

An investigation of \ac{HO} timing provides further insights into \ac{HO} performance.
Fig.~\ref{fig:ecdf} shows that the \ac{HO} timing achieved by the \ac{DRL}~agent is better than the timing of the 3GPP reference.
The figure displays the \ac{ECDF} of the \ac{SINR} at the start (solid lines) and end (dashed lines) of the \ac{HO} execution, representing the \ac{SINR} of the serving \ac{BS} upon receipt of the \ac{HO} command and the \ac{SINR} of the target \ac{BS} at the \ac{HO} completion.
Additionally, \ac{QoS} thresholds for $Q_\text{out}$ and $Q_\text{in}$, important for \ac{RLF} detection, are marked as solid/dashed red lines.
The results show that the probability $P(\text{SINR} \leq Q_\text{out})$ at the time of \ac{HO} is reduced by more than 30\% compared to the reference, leading to a decrease in \acp{HOF}, as shown in Fig.~\ref{fig:rlf-pp-number}.

It is also worth mentioning that the \ac{SINR} of the \ac{PPO}~agent post-\ac{HO} completion is, on average, slightly lower than that of the reference protocol, due to the agent’s earlier \ac{HO} timing.
However, the difference in probabilities within the \ac{RLF}-relevant \ac{SINR} range is negligible, underscoring the improved \ac{HO} timing. The lower \ac{HOF} probability in Fig.~\ref{fig:rlf-pp-number} confirms the effectiveness in reducing the \acp{RLF} frequency caused by \acp{HO}.

Fig.~\ref{fig:rlf-pp-number} compares the average \ac{HOF} and \ac{PP} probability of the 3GPP protocol and the \ac{PPO} agent for different \ac{UE} speeds. The \ac{PPO} agent outperforms the 3GPP protocol in terms of \ac{HOF} probability, while the \ac{PP} probability is at a similar level for both protocols.
The \ac{PP} rate of the 3GPP protocol is relatively high due to the chosen parameter \ac{TTT} of $\SI{40}{ms}$, but leads to a higher average rate than larger \acp{TTT}.

\pgfplotsset{scaled y ticks=false}
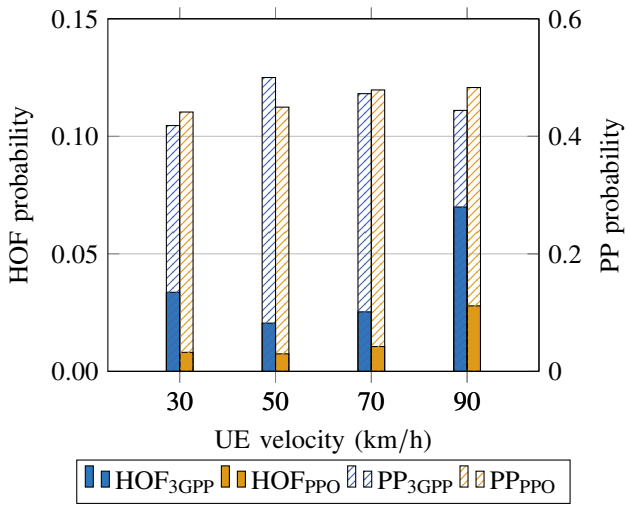
\begin{figure}[ht]
    \centering
    \begin{tikzpicture}
    \begin{axis}[
        ybar=0pt,
        bar width=5pt,
        symbolic x coords={20,30, 50, 70, 90},
        xtick=data,
        ymin=0,
        ymax=0.15,
        ymajorgrids=true,
        ylabel={HOF probability},
        xlabel={UE velocity ($\SI{}{km/h}$}),
        legend style={at={(0.5,-0.15)}, anchor=north, legend columns=-1},
        enlarge x limits=0.25,
        legend cell align={left},
        yticklabel style={
            /pgf/number format/fixed,
            /pgf/number format/precision=2,
            /pgf/number format/fixed zerofill
        },
        scaled y ticks=false,
    ]
    \addplot[
        ybar,
        fill=kit-blue100
    ] coordinates {
        (30,0.03365384615384615)
        (50,0.020522388059701493)
        (70,0.02531645569620253)
        (90,0.06993006993006994)
    };
    \label{leg3gpp}
    \addplot[
        ybar,
        fill=KITorange
    ] coordinates {
        (30,0.008097165991902834)
        (50,0.007462686567164179)
        (70,0.010563380281690139)
        (90,0.027863777089783284)
    };
    \label{legppo}
    \end{axis}
    
    \begin{axis}[
        ybar=0pt,
        bar width=5pt,
        symbolic x coords={20,30, 50, 70, 90},
        xtick=data,
        ymin=0,
        ymax=0.6,
        axis y line*=right,
        ylabel={PP probability},
        yticklabel style={anchor=west},
        yticklabel pos=right,
        enlarge x limits=0.25,
        legend style={at={(0.5,-0.25)}, anchor=north, legend columns=-1}
    ]
    \addlegendimage{/pgfplots/refstyle=leg3gpp}\addlegendentry{$\text{HOF}_\text{3GPP}$}
    \addlegendimage{/pgfplots/refstyle=legppo}\addlegendentry{$\text{HOF}_\text{PPO}$}
    \addplot[
        ybar,
        pattern=north east lines,
        pattern color=KITblue
    ] coordinates {
        (30,0.4182692307692308)
        (50,0.500)
        (70,0.4725738396624472)
        (90,0.444055944055944)
    };
    \addplot[
        ybar,
        pattern=north east lines,
        pattern color=KITorange
    ] coordinates {
        (30,0.44129554655870445)
        (50,0.4496268656716418)
        (70,0.47887323943661964)
        (90,0.48297213622291024)
    };    
    \addlegendentry{$\text{PP}_\text{3GPP}$}
    \addlegendentry{$\text{PP}_\text{PPO}$}
    \end{axis}

    \end{tikzpicture}
    \caption{\ac{HOF} and \ac{PP} probability.}
    \label{fig:rlf-pp-number}
\end{figure}

    \section{Conclusion}\label{sec:6_conclusion}
This paper has presented a reinforcement learning-based approach using \acl{PPO} to enhance handover protocols in dense cellular networks.
By focusing on optimal handover timing and decisions, the proposed \ac{PPO}-based agent adapts dynamically to various user mobility patterns and network conditions.
Our results demonstrate that the adaptive approach achieves at least the same average data rates, but reduces \acp{RLF} compared to the 3GPP-standardized 5G NR \ac{HO} protocol, especially in high-mobility scenarios.
The adaptability of \ac{DRL}-based protocols offers promising potential for next-generation cellular networks towards more reliable and efficient connectivity in \acp{HetNet}.
This work contributes to the growing field of mobility management optimization, with publicly available source code and datasets for further research and development.

    \bibliographystyle{IEEEtran}
    \bibliography{IEEEabrv,references}

    \begin{acronym}
        \acro{GAE}{generalized advantage estimation}
        \acro{ECDF}{empirical cumulative distribution function}
        \acro{BS}{base station}
        \acro{UE}{user equipment}
        \acro{SUMO}{Simulation of Urban Mobility}
        \acro{RSRP}{Reference Signal Received Power}
        \acro{SINR}{signal-to-interference-plus-noise ratio}
        \acro{TTT}{time-to-trigger}
        \acro{HO}{handover}
        \acro{RLF}{radio link failure}
        \acro{HOF}{handover failure}
        \acro{PP}{ping-pong}
        \acro{RL}{reinforcement learning}
        \acro{DRL}{deep reinforcement learning}
        \acro{PPO}{proximal policy optimization}
        \acro{MTS}{minimum-time-of-stay}
        \acro{NN}{neural network}
        \acro{QoS}{quality of service}
        \acro{LTE}{LTE}
        \acro{HetNet}{heterogeneous network}
    \end{acronym}
\end{document}